\documentclass[prb,aps,twocolumn,showpacs,citeautoscript,longbibliography]{revtex4-1}

\usepackage{amsmath}
\usepackage{bm}
\usepackage{amssymb}
\usepackage{graphicx}
\usepackage[colorlinks=true,allcolors=blue]{hyperref}

\begin{document}

\title{Korshunov instantons in a superconductor at elevated bias current}

\author{Daniel Otten}
\affiliation{JARA-Institute for Quantum Information, RWTH Aachen University,
D-52056 Aachen, Germany}

\author{Fabian Hassler}
\affiliation{JARA-Institute for Quantum Information, RWTH Aachen University,
D-52056 Aachen, Germany}


\begin{abstract} 
  Even at zero temperature dissipation reduces quantum fluctuations and tends
  to localize particles. A notable exception is the nonlinear dissipation due
  to quasiparticle tunneling in a Josephson junction. It is well known that
  quasiparticle dissipation does not suppress tunneling of the superconducting phase difference between next-nearest metastable phase states even though tunneling to the nearest phase state is suppressed. The reason is
  that the dissipative action admits an instanton solution, the so-called
  Korshunov instanton. Here, we analyze this model at elevated bias current
  $I$. We find that besides the known regime where the logarithm of the
  tunneling rate scales as $I^{2/3}$ there is novel regime with a scaling
  $I^2$. We argue that the increased tunneling rate that derives from the
  elevated bias current is favorable for  an experimental verification of the
  Korshunov instantons.
\end{abstract}

\maketitle
\section{Introduction}
Dissipative effects in quantum mechanics have been an object of research for a
long time. With their simple model of quantum dissipation, Caldeira and
Leggett popularized a linear and analytically trackable model to effectively
describe dissipation in quantum systems \cite{Caldeira1981,Weiss2012}. Despite
the tremendous success of this simple model, it cannot describe all
dissipative effects; for example, it does not describe quantization of the
transported charge in a resistor built from a tunnel junction. To this end, the
Ambegaokar-Eckern-Sch\"on (AES) model was introduced modeling a
superconducting tunnel junction that is subject to quasiparticle tunneling
\cite{Ambegaokar1982}. The latter introduces a dissipative term in the action
that is periodic in the superconducting phase difference with a period of
$4\pi$ corresponding to the normal flux quantum. In
Ref.~\onlinecite{Guinea1986}, it was noted that due to quasiparticle tunneling the ground state is either a
superposition of the phase localized in the even or odd minima of the
Josephson potential.  Remarkably, these superpositions are immune to the
quasiparticle dissipation and survive even in the limit of strong dissipation.
As a consequence, the phase `particle' does not localize in one of the
minima \cite{Korshunov1987}.  The action describing the quasiparticle
tunneling admits instanton solutions, the so-called Korshunov instantons, that
connect minima in the Josephson potential separated by $4\pi$.  This results
in a coherent tunneling amplitude between states localized in next-nearest
minima and, consequently, the formation of bands. However, the band-width is
exponentially small in the damping parameter and thus the experimental
verification remains challenging.

In a different context, Korshunov instantons are important to understand
charging effects of metallic islands connected to reservoirs via tunnel
junctions \cite{Panyukov1991,Nazarov1999,Altland2006,bollinger2006,Titov2016}.
Recently, the charging energy of a normal island has been measured as a
function of the tunnel coupling \cite{Jezouin2016}.  However, to our
knowledge, a direct observation of Korshunov instantons of the superconducting
phase tunneling to the next-nearest minima is still missing. This observation
would not only be interesting as an example where strong dissipation does not
completely suppress tunneling but also because Korshunov instantons are
related to coherent, paired phase slips which are of interest for example for
the realization of parity protected
qubits\cite{doucot2002,ioffe2002,gladchenko2009, bell2014}.  Additionally, the
system is an interesting example of dissipative quantum mechanics with a
multitude of different regimes that can be accessed by simply changing the
bias current; the regimes cover coherent quantum dynamics, even in the
presence of strong dissipation, the case of a special incoherent relaxation
due to quasiparticle tunneling, and more conventional Ohmic relaxation
\cite{Korshunov19872,Korshunov1987}.

In this work, we investigate the effect of an elevated bias current $I$ on
Korshunov instantons at zero temperature. This is relevant because increasing
the bias current raises the tunneling rate $\Gamma_{4\pi}$ of the
superconducting phase to the next-nearest minima of the potential that is the
hallmark of the presence of strong quasiparticle dissipation and thus
increases the chance of experimental verification of the theoretical results.
We find that, apart from the low bias regime with $\ln(\Gamma_{4\pi})\propto
I^{2/3}$,  studied in Ref.~\onlinecite{Korshunov1987}, there is a novel regime
at elevated bias current where $\ln(\Gamma_{4\pi})\propto I^{2}$. At even
higher bias current, the quasiparticle nature of the dissipation becomes
irrelevant and only tunneling to the next-nearest minima survives that is
described by the conventional Ohmic model of Caldeira and Leggett. We discuss
on the transition between the different regimes and propose an experimental
method to measure the predicted decay rates.

The paper is organized as follows. In Sec.~\ref{Sec.:Setup}, we introduce the
setup and the theoretical model. In \ref{Sec.:InstantonAnalysis}, we provide a
short introduction to the notation and the instanton method that we use
throughout this work. In \ref{Sec.:CoherentTunneling}, we give a comprehensive
derivation of the coherent tunneling amplitude before we proceed with the
calculation of the incoherent tunneling rates in
Sec.~\ref{Sec.:IncoherentTunneling}. Note that these sections have some
overlap with the work of Ref.~\onlinecite{Korshunov1987}.
Section~\ref{Sec.:IncoherentTunneling} includes our main result of the scaling
of the tunneling rate at elevated bias current. Moreover, we discuss the
transition between the coherent, incoherent, and conventional Ohmic regime.
In \ref{Sec.:Measurement}, we propose a simple scheme on how to measure the
incoherent tunneling rate before we end with our conclusions.

\section{Setup} \label{Sec.:Setup}

For our analysis, we consider a current biased tunnel junction between two
superconducting leads that is intrinsically subject to quasiparticle tunneling
that acts as a  dissipative element. This can be described by the AES model.
In the Euclidean (imaginary time) path-integral formalism, its dimensionless
action $S=S_c+S_\eta$ at zero temperature is given  by\cite{Ambegaokar1982}
\begin{align}
\label{Eq:Action}
S_c &=\int_{-\infty}^{\infty} \!dt \Biggl[\frac{\hbar C}{8 e^2}\dot{\varphi}^2-\frac{E_J}{\hbar} [1-\cos(\varphi)]+\frac{I\phi_0}{\hbar}\varphi\Biggr],\nonumber\\ 
S_\eta &=\frac{\hbar}{\pi e^2R}\int_{-\infty}^{\infty}\! dt\, dt' \frac{\sin\{[\varphi(t)-\varphi(t')]/4\}^2}{(t-t')^2},
\end{align}
where $\varphi$ is the superconducting phase difference across the Josephson
junction and $\dot{\varphi}=d\varphi/dt$ its derivative with respect to the
imaginary time $t$. The first  term $S_c$ describes the coherent
superconducting circuit consisting of a Josephson junction with Josephson
energy $E_J=\phi_0 I_c/2\pi$, where $I_c$ is the junctions critical current
and $\phi_0=2e/h$ the superconducting flux quantum. The capacitive energy due
to the junctions capacitance $C$ is given by $E_C=e^2/2C$. The second term
$S_\eta$ (quasiparticle action) corresponds to the dissipation due to
quasiparticle tunneling. Its magnitude is connected to the effective shunt
resistor $R$. For small $\varphi$, the dissipative action can be expanded in
a Taylor series so that it reproduces the Ohmic action described by Caldeira
and Leggett. However, this approach neglects the periodicity of $S_\eta$. The
latter causes the action to stay invariant for a $4\pi$-phase shift
corresponding to the tunneling of a normal flux quantum \cite{ulrich2016}. This
refers to the fact that the quasiparticles are quantized single electrons and
therefore do not feel a shift of a normal flux quantum.

In this work, we are interested in the regime where the dissipative action
$S_\eta$ dominates $S_c$ with $\hbar/4e^2R\gg (E_J/8E_C)^{1/2}$. Such a strong
dissipation brings the system always into the semiclassical regime so that an
instanton analysis is applicable. Interestingly, the quasiparticle action by
itself can admit instanton saddle points without an additional kinetic or
potential term.  Therefore, the solution of $\delta S_\eta/\delta\varphi=0$,
where $\delta S_\eta/\delta\varphi$ is the first variation of the
quasiparticle action, is an approximative saddle point of the full action $S$.
In Ref.~\onlinecite{Korshunov1987}, it was shown that there exists a solution
for this equation, the Korshunov instanton $\varphi_I(t)=4 \arctan(\Omega t)$,
with arbitrary frequency $\Omega$, that connects not neighboring minima of the
Josephson potential but next-nearest minima. For vanishing bias current $I=0$,
it is this instanton of the dissipative action that results in a coherent
tunnel element between minima shifted by $4\pi$ and leads to the formation of
bands even in the presence of strong dissipation. However, the resulting
bandwidth is small and difficult to tune and the effect of the pure coherent
tunneling therefore is difficult to measure. The situation can be changed by
applying a bias current $I$. On one hand, this destroys the bands but on the
other hand it introduces a dissipative incoherent tunneling rate where a phase
particle  located in one of the minima tunnels by $4\pi$ to the next-nearest
minimum. Additionally, the bias gives rise to `Ohmic' decay into the next
minimum for which the quasiparticle action acts as a simple Ohmic shunt.
Contrary to intuition, at low bias  current, the $4\pi$ tunneling dominates
the $2\pi$ tunneling, \textit{i.e.}, the particle is more likely to tunnel to
the next-nearest minimum than to the nearest minimum. While the $2\pi$
tunneling vanishes at zero bias, the $4\pi$ process transforms into the
coherent tunneling element. 

For the analysis, we introduce the dimensionless parameters
\begin{align}
	j=I\phi_0/E_J,\;\eta=\hbar/4e^2R\,\; \text{and} \;
	\zeta=(E_J/8E_C)^{1/2}.
\end{align}   
The normalized bias current $j$ gives a measure of how strong the potential is
tilted. For $j=1$, the tilt due to the bias is so strong that the minima in
the potential vanishes. At this point, the particle classically slides down
the potential landscape. The parameter $\zeta$ describes the ratio between the
capacitive kinetic energy and the Josephson potential energy. Without
dissipation, it describes the quantum uncertainty of the phase with $\delta
\phi \propto \zeta^{-1}$. The parameter $\eta$ describes the strength of
dissipation. For large $\eta$, the dissipation is strong and the phase becomes
localized. Note that for $\eta\gg1$, semi-classical methods are applicable
even for $\zeta<1$.

\section{Saddle Point Approximation} \label{Sec.:InstantonAnalysis} 

In this section, we concisely describe the instanton method for analyzing
tunneling problems. In the following sections, coherent tunnel elements as
well as incoherent tunneling rates will be calculated. Both can be
accomplished by evaluating the imaginary time path integral in Gaussian
approximation around a saddle point $\bar{\varphi}(t)$ of the action $S$. In
general, the action admits different saddle points with different physical
meanings. Given a saddle point, the imaginary time propagator can be
approximated as
\begin{align}
\label{Eq.:ImaginaryTimeProp}
G[\bar{\varphi}(t)]=\int_{\varphi\approx \bar{\varphi}}\!\!
\mathcal{D}[\varphi]e^{-S_\mathcal{G}}.
\end{align}
Here, $\bar{\varphi}$ is defined as the solution of $\delta S/\delta\varphi
=0$ with appropriate boundary conditions,
 $\mathcal{D}[\varphi]$ is the functional integration measure, while the
subscript $\varphi\approx\bar{\varphi}$ indicates that the path integral
should be evaluated in Gaussian approximation around the extremum
$\bar{\varphi}$.

The action $S_\mathcal{G}$ corresponds to $S$
expanded to second order in the fluctuations deviating from the extremal
path.  In particular, we set
\begin{align}
\label{Eq.:ExpandedPhase} \varphi(t)=\bar{\varphi}(t)+\sum_n c_n \chi_n(t),
\end{align}
with $n\in\mathbb{N}_0$. The approximated action
$S_\mathcal{G}$ can be written as
\begin{align}
  S_\mathcal{G}&=S_{\bar{\varphi}}+ \sum_{n,n'} c_n c_{n'} \int \!dt\,
\chi_n\frac{\delta^{2} S}{\delta^{2}\varphi}
[\bar{\varphi}]\,\chi_{n'}\nonumber\\ &
=S_{\bar{\varphi}}+\sum_n\Lambda_n c_n^2,
\end{align}
where $S_{\bar{\varphi}}$ is the action directly evaluated at the extremal
path $\bar{\varphi}$. For the second equality, we have assumed that the
fluctuation modes $\chi_n$ are eigenfunctions to the
second variation satisfying
\begin{align}
\label{Eq.:EigenvalueFluctuations} 
\frac{\delta^2S}{\delta\varphi^2}
[\bar{\varphi}]\,\chi_n=\Lambda_n\chi_n 
\end{align}
with eigenvalues $\Lambda_n$ and normalized to  $\int\! dt\,
\chi_{n}(t)\chi_{n'}(t)=\delta_{n,n'}$. With this, the integration measure can
be chosen to be $\mathcal{D}[\varphi]=\mathcal{N} \prod_n dc_n$ where
$\mathcal{N}$ is a normalization constant. Every positive $\Lambda_n$ leads to
a Gaussian integral with the result
\begin{align} \label{Eq.:ResultingImaginaryTimeProp}
  G[\bar{\varphi}]&=\mathcal{N} \int \prod_n dc_n
\exp[-S_{\bar{\varphi}}-\sum_n\Lambda_n c_n^2]\nonumber\\ 
&=\mathcal{N} \prod_n
(\pi/\Lambda_n)^{-1/2}e^{-S_{\bar{\varphi}}}\nonumber\\
&=Fe^{-S_{\bar{\varphi}}}.
\end{align} 
For an instanton solution $\bar{\varphi}$, we have to deal with a zero
eigenvalue that cannot be treated by the simple Gaussian integration above.
Handling it correctly\cite{Coleman1979,larkin1984} leads to the prefactor
$F=\omega_0A_1 A_2$ with\cite{Weiss2012}
\begin{align}
\label{Eq.:AFactorsGeneral1} A_{1} &= \sqrt{\frac{W}{2 \pi}}
\frac{\hbar\omega_0^{2}}{8E_C\sqrt{\Lambda_1 \Lambda_2}}\\
\label{Eq.:AFactorsGeneral2} A_{2} &= \frac{8E_C\prod_{n=1}
^{\infty}\Lambda_{n,0}^{1/2}}{\hbar\omega_0 ^{2} \prod_{n=3} ^{\infty}
\Lambda_n^{1/2}} \end{align} 
Here, the frequency $\omega_0=(8E_J E_C)^{1/2}/\hbar$ denotes the plasma
frequency and the factor $A_1$ incorporates the product of the three lowest
eigenvalues including the zero eigenvalue. The zero mode is accounted for by
the expression $W=\hbar\int \!dt \,\dot{\bar{\varphi}}^2/8E_C$. The factor
$A_2$ includes the eigenvalues $\Lambda_n$ with $n\geq 3$. Its leading
behavior is determined by the asymptotics for $n\rightarrow \infty$. The
$\Lambda_{n,0}$ correspond to the fluctuations around the constant path
$\varphi_0=0$. They enter the equation when fixing the normalization
$\mathcal{N}$.

To conclude this section, we shortly discuss the applicability of the
semiclassical approximation above. It corresponds to the method of steepest
decent that is applicable as long as $S_{\bar \varphi}$ is much larger than
one.  Additionally, within one potential well, the phase should be localized
in the minimum. While this condition normally demands $E_J\gg E_C$, it is
always fulfilled in the case of strong dissipation $\eta \gg 1$ as the
dissipation localizes the phase difference across the Josephson junction.

\section{Coherent Tunneling} \label{Sec.:CoherentTunneling}

\begin{figure}
\includegraphics[width=.9\linewidth]{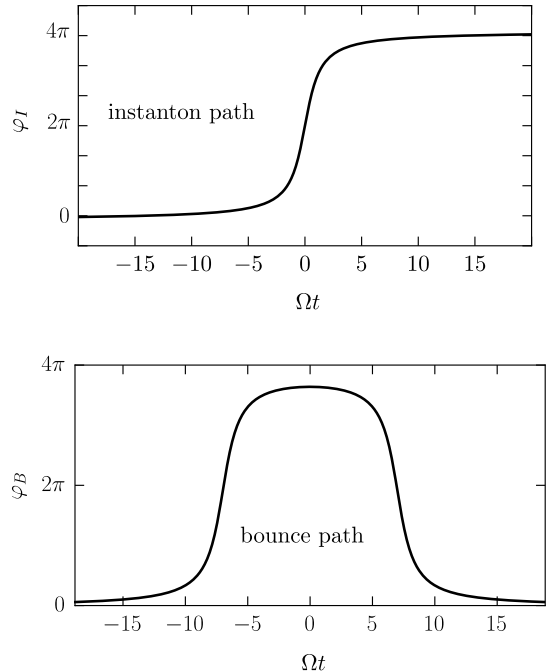}
\caption{\label{Fig:paths}The upper panel shows the instanton path $\varphi_I$
connecting the minima of the Josephson potential at $\varphi=0$ and at
$\varphi=4\pi$. The instanton solution corresponds to coherent tunneling of
the phase difference. The lower panel shows the bounce path $\varphi_B$, a
closed trajectory connecting the origin with itself via a fast penetration of
the potential barrier. In our case, it consist of a superposition of an
instanton shifted by $\tau/2$ in imaginary time with an anti-instanton shifted
by $-\tau/2$. For the plot, we have chosen a value $\Omega\tau=20$. It is
related to the incoherent decay out of the potential minimum at the origin,
see main text in Sec.~\ref{Sec.:IncoherentTunneling}.}
\end{figure}
Coherent quantum tunneling describes the Hamiltonian evolution of a system
that connects localized states separated by a classically inaccessible
barrier. This unitary evolution leads to quantum superposition of the particle
in different potential wells. In our case, the system is mainly localized in
the minima of the Josephson potential, \text{i.e.}, at $\phi\in
2\pi\mathbb{Z}$.  This makes it possible to treat the minima of the cosine
potential as sites of a linear lattice. The tunneling between different sites
causes the formation of bands with a bandwidth $\Delta_I$ equivalent to twice
the tunneling matrix element. The bandwidth can be expressed by the imaginary
time propagator evaluated at the so-called instanton $\varphi_I$. It is a
saddle point of the action connecting two minima of the Josephson potential.
It can be shown that the bandwidth is given by $\Delta_I=4\hbar
G[\varphi_I]=4\hbar F_Ie^{-S_I}$,  where $S_I$ is the action evaluated at the
instanton saddle point $\varphi_I$ and $F_I$ originates from the Gaussian
fluctuations around this instanton path \cite{Coleman1979}.

\subsection{Instanton Action}

We are going to determine the extremal action corresponding to an instanton
that connects two minima of the Josephson potential. For this analysis, we are
essentially following Ref.~\onlinecite{Korshunov1987}. The saddle point
equation $\delta S_\eta/\delta\varphi=0$ reads
\begin{align}
\label{Eq.:FirstOrderVariation}
\frac{\delta S_\eta}{\delta\varphi}[\varphi_I]=\frac{2\eta}{\pi} \int\! dt'\, 
\frac{\sin\{[\varphi(t)-\varphi(t')]/2\}}{(t-t')^{2}}=0.
\end{align}
An instanton solution to this equation is given by \cite{Korshunov1987}
\begin{align}\label{eq:inst}
  \varphi_I(t)=4 \arctan[\Omega (t-\tau) ],
\end{align}
connecting a minimum of the $\cos$-potential at $t=-\infty$ with a next to
nearest neighbor minimum shifted by $4\pi$ at $t=\infty$. It depends on the
frequency $\Omega$ that determines how fast the phase flips. The solution
$\varphi_I$ is in principle only a saddle point of the quasiparticle action
$S_\eta$ and not of the full action $S$.  However, in the case $\eta \gg
\zeta$, the quasiparticle action dominates the saddle point solution and thus
even including the circuit action $S_c$ in Eq.~(\ref{Eq.:FirstOrderVariation})
changes the instanton only perturbatively.  Therefore, it is justified to
insert the quasiparticle instanton $\varphi_I$ into the action $S_c$ which
corresponds to proceeding with first order perturbation theory. We find as
resulting action $S_I (\Omega)$ on the instanton path
\begin{align}
S_I (\Omega)=4 \pi \biggl(\eta + \frac{\hbar \Omega}{8E_C}  + \frac{E_J}{\hbar \Omega}\biggr).
\end{align}
The action depends on $\Omega$ so that we also need to extremize with respect to this
parameter. We find a minimum of the action where $\Omega$ is equal to the
plasma frequency $\omega_0$ of the minimum with $\Omega=\omega_0=(8E_J E_C)^{1/2}/\hbar$ . At this minimum the instanton action becomes 
\begin{align}
S_I =4 \pi (\eta + 2\zeta).
\end{align}

\subsection{Instanton Prefactor and Result}

The next step is the evaluation of the fluctuations to determine the prefactor
$F_I$. The explicit action of the fluctuation operator on the $\chi_n$ is
given by
\begin{align}
\label{Eq.:CircuitFluctuations}
\frac{\delta^2 S_c}{\delta\varphi^2}[\varphi_I]\,\chi_n(t)
=&\biggl[-\frac{\hbar}{8E_C} \frac{\partial^2 }{\partial t^2}+\frac{E_J}{\hbar}\cos(\varphi_I)\biggr]\chi_n(t), \\
\label{Eq.:DissiFluctuations}
\frac{\delta^{2} S_\eta}{\delta^{2}\varphi} [\varphi_I] \,\chi_n(t) =&
\frac{\eta}{\pi} \int \!dt'\, \frac{\cos\{[\varphi_I (t) - \varphi_I (t')]/2\}}{(t-t')^{2}}\nonumber\\ 
&\hspace{50pt}\times [\chi_n (t)- \chi_n (t')],
\end{align}
where we separated the operator in circuit and dissipative contributions. By
acting on the $\chi_n$, these operators define a stationary Schr\"odinger
equation with a non-local potential. Here, the imaginary time plays the role
of the spatial coordinate. The lower eigenvalues are mainly determined by the
dissipative action corresponding to bounded states in the potential. However,
for the high energy modes, it is the kinetic energy term that dominates and
gives rise to a continuum of states lying above the bounded spectrum. 
For ease of mode counting, we temporarily introduce a finite imaginary time
interval $\beta$ with periodic boundary conditions, corresponding to
nonzero temperatures. At the end, we send the interval to infinity again.  

For low energies, only the dissipative action is relevant. The eigenvalue
equation related to Eq.~(\ref{Eq.:DissiFluctuations}) is given explicitly as
(for $\tau =0$)
\begin{multline}
\label{Eq.:FluctuationDissipation}
\frac{-2 \Omega}{1+(\Omega t)^{2}}
\biggl[ \chi_n (t) 
  -\int \!\frac{dt'}{\pi}
\frac{\Omega\,\chi_n
(t')}{1+ (\Omega t')^{2}} \biggr] \\
+  \int\!\mathcal{P} \frac{dt'}{\pi} \frac{1}{(t-t')}
 \frac{d \chi_n(t')}{d t'} 
=\frac{\Lambda_{n}}{\eta} \chi_n (t),
\end{multline}
where the $\mathcal{P}$ denotes the Cauchy principle value. In general such an
equation is hard to solve. However, we obtain a zero mode for each free
parameter of the instanton solution Eq.~\eqref{eq:inst}, which are in our case
the imaginary time $\tau$ and the frequency $\Omega$. These zero modes generate a shift or dilation of the solution in imaginary time without changing the value
of the action $S_\eta$.  The zero modes can be found by taking the derivative
of the instanton path with respect to the corresponding free parameters. We
find
\begin{align}
  \chi_0&= N_0 \frac{d\varphi_I(t)}{d\tau}=\sqrt{\frac{2}{\pi}}
  \frac{\Omega^{1/2}}{1+ (\Omega t)^{2}},\nonumber\\
\chi_1&=N_1 \frac{d\varphi_I(t)}{d\Omega}=\sqrt{\frac{2}{\pi}}
\frac{\Omega^{3/2}t}{1+ (\Omega t)^{2}},
\end{align}
both with eigenvalue $\Lambda_0=\Lambda_1=0$; the normalization $N_j$ fixed by
$N_{j}^2\int \!dt\, \chi_j^2=1$.

It is well known that for Schr\"odinger like equations the number of nodes in
the eigenfunction can be associated with the size of the eigenvalue, where the
eigenfunction with the lowest number of nodes corresponds to the lowest
eigenvalue \cite{landau:3}. For higher modes, the zero modes should be
modulated in order to obtain more nodes \cite{larkin1984}. For $n=2$, we
obtain approximately \cite{Korshunov1987}
\begin{align}
\chi_{2}=&\biggl(\frac{2}{\beta}\biggr)^{1/2}\frac{\cos(\nu_{1}
t)(1-t^2\Omega^2)+\sin(\nu_{1} t)2t\Omega}{1+t^2\Omega^2}
\end{align} 
with the eigenvalue $\Lambda_{2}=\eta\nu_{1}$, where $\nu_n=2\pi n/\beta$ are
the bosonic Matsubara frequencies.  We incorporate the effect of $S_c$ by
performing lowest-order perturbation theory with
\begin{align}
\label{Eq.:InstantonPerturbation1}
\Lambda_n=\int \!dt\, \chi_n\,\frac{\delta^2 S}{\delta\varphi^2}[\varphi_I]
\,\chi_n.
\end{align}
We obtain $\Lambda_0=0$,  $\Lambda_1= \hbar\Omega^2/16E_C$, and
$\Lambda_2=E_J/\hbar=\hbar\Omega^2/8E_C$ for the lowest three eigenvalues determining
$A_1$.

The calculation of $A_2$ we first consider only the kinetic term in
(\ref{Eq.:CircuitFluctuations}) and treat the rest as a perturbation. The
eigenfunctions of the kinetic operator are given by
$\chi_{2n}=(2/\beta)^{1/2}\sin(\nu_{n} t)$ and
$\chi_{2n+1}=(2/\beta)^{1/2}\cos(\nu_{n} t)$ with eigenvalues
$\Lambda_{2n}=\Lambda_{2n+1}=\hbar\nu_{n}^2/8E_C$. By treating the rest of the
action in first order perturbation theory, the eigenvalues at large $n$ are
given by
\begin{align}
\label{Eq.:UpperNEigenvalues}
\Lambda_{2n-1}=\Lambda_{2n}=&\int \!dt \,\chi_n\frac{\delta^2
S}{\delta\varphi^2}[\varphi_I]\,\chi_n\nonumber\\
=&\frac{\hbar}{8 E_C}\bigl(\nu_n^2+\omega_0^2\bigr)+\eta|\nu_n|-\eta\nu_1.
\end{align}
The term proportional to $\omega_0^2$ originates from the fluctuations in the
Josephson potential while $\eta|\nu|$ is produced by the last term in
(\ref{Eq.:FluctuationDissipation}). The $n$-independent offset $\eta\nu_1$ is
generated by the first part of the first term in
(\ref{Eq.:FluctuationDissipation}), whereas the integral without the principal
part does not contribute for large $n$ because it is exponentially suppressed
by the factor $e^{-\nu_n}$. For the normalization of $A_2$ we also need the
eigenvalues $\Lambda_{n,0}$ corresponding to the fluctuations around the
constant path $\varphi_0=0$. These are given by
\begin{align}
 \Lambda_{2n-1,0}=\Lambda_{2n,0}=\frac{\hbar}{8 E_C}\bigl(\nu_n^2+\omega_0^2\bigr)+\eta|\nu_n|
\end{align}
and correspond to $\Lambda_n$ in Eq.~(\ref{Eq.:UpperNEigenvalues}) without the
offset $\eta \nu_1$.

With the eigenvalues at hand we are in the position to evaluate $A_2$.
Evaluating the infinite product ratio (\ref{Eq.:AFactorsGeneral2}) we can
write in our regime $\eta\gg\zeta$
\begin{align}
\label{Eq.:A2Gammas}
A_2=\frac{\eta^2}{\zeta^2}.
\end{align} 
Using the results \eqref{Eq.:InstantonPerturbation1} in
\eqref{Eq.:AFactorsGeneral1}, \eqref{Eq.:A2Gammas}, and the zero mode
normalization $W_I=\pi \hbar\Omega/E_C$, the final expression for the
bandwidth is given by\cite{korshunov}
\begin{align}
  \Delta_I=4\frac{\eta^2\hbar\Omega}{\zeta^{3/2}}e^{-4\pi(\eta+2\zeta)}.
\end{align}

\section{Incoherent tunneling }
\label{Sec.:IncoherentTunneling}

\begin{figure}
\centering
\includegraphics[width=.9\linewidth]{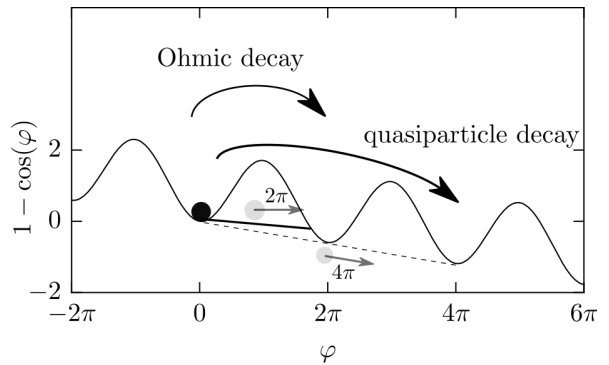}
\caption{\label{Fig:Pot}The plot shows the Josephson potential biased by a
current $j=0.1$. The black circle corresponds to the phase difference localized at the
origin. Conventional Ohmic decay tunnels the phase through the potential
barrier to the next minimum as shown by the black line. The decay due to the
quasiparticle tunneling is only slightly influenced by the potential and
directly goes to the minimum shifted by $4\pi$, indicated by the dashed line.}
\end{figure}
Switching to a finite bias current $j$, we render the minima in the Josephson
potential unstable. Considering the Hamiltonian time evolution in this system,
we cannot treat the minima of the Josephson potential as sites with a single
level of a tight binding model as for the case of coherent tunneling. The
evolution brings the initial state into a superposition of excited states of
the neighboring minimum. Only the strong dissipation then localizes these
state again in the local minimum. Such an evolution is called incoherent
tunneling.  For intermediate evolution times, this can be approximated as an
exponential relaxation out of the original well and can be expressed by an
imaginary part of the energy when starting in a single minimum. For this
problem, the important object is not the instanton trajectory but the bounce
$\varphi_B$.  This is a cyclic trajectory connecting the minimum with a
turning point and going back to its starting point as shown in the lower plot
of Fig.~\ref{Fig:paths}. It can be shown\cite{Coleman1979} that in this case,
the incoherent decay rate $\Gamma_{4\pi}$ is given by
$G[\varphi_B]=F_Be^{-S_B}$; here, $S_B$ is the action $S$ evaluated at the
bounce trajectory and in \eqref{Eq.:AFactorsGeneral1} we have to replace
$\Lambda_1$ by $|\Lambda_0|$ because of an occurring negative eigenvalue of the second variation, see below.

\subsection{Bounce Action}\label{sec:bounce}
\begin{figure}
 \centering 
\includegraphics[width=.9\linewidth]{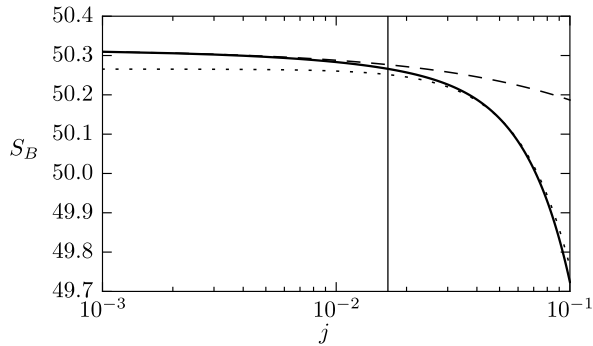}
\caption{\label{Fig:ActionPlots} The plot shows the value of the bounce action  $S_B(\Omega,\tau)$ at the saddle point plotted versus the bias current $j$ for $\eta=2$ and $\zeta=10^{-3}$. The solid black line corresponds to the saddle point obtained numerically, while the dashed line corresponds to $S_B^{(\text{i})}$ (Eq. (\ref{Eq.:KorshunovAction})) with a $j^{2/3}$-dependence. The dotted line depicts the action $S_B^{\text{(ii)}}$ (Eq. (\ref{Eq.:BounceActionResult})) with a $j^2$-dependence. The solid vertical line marks the crossover between the two regimes (i) and (ii) at $j=(\zeta/2\eta)^{1/2}\approx 0.016$. We observe that the validity of the solution $S_B^{(i)}$ breaks down for elevated bias currents and the action changes its behavior from a $j^{2/3}$-dependence to a $j^2$-dependence.}
\end{figure}
We start this section with the discussion of the bounce action $S_B$. In
principle, as the quasiparticle action dominates, it is justified to find a
saddle point of only the quasiparticle action and treat the circuit action in
perturbation theory, as in the case of the instanton. However, there is also a
bounce solution that is mainly determined by the circuit action. It
corresponds to the tunneling of the phase difference through the barrier between
the origin and the nearest neighbor minimum of the Josephson potential. In
Fig.~\ref{Fig:Pot}, this is indicated by the arrow labeled with $2\pi$. For
such a trajectory, we can expand the quasiparticle action to second order so
that it reproduces conventional Ohmic dissipation. Therefore, we call the
decay due to this bounce solution in the following `Ohmic decay'. It results
at low temperatures in the decay rate $\Gamma_{2\pi}\propto
j^{4\pi\eta-1}$\cite{Korshunov19872}. For small currents, this rate is lower
than the rate of decay to the next-nearest minimum caused by the quasiparticle
action. While it accounts for a $2\pi$ phase slip, the quasiparticle bounce
corresponds to a paired $4\pi$ phase slip into the next-nearest minimum.
Therefore, both processes can physically be distinguished and should be
individually considered. In the following, we calculate the dominating rate of
decay due to the quasiparticle tunneling.

The analytical solution to the saddle point problem of $S_\eta$ that fulfills
the boundary conditions of the bounce is not known. However, we can construct
an asymptotic saddle point by adding an instanton shifted by $\tau/2$ in
imaginary time with an anti-instanton shifted by $\tau/2$ in the other
direction resulting in the bounce path
$\varphi_B=\varphi_I(t+\tau/2)-\varphi_I(t-\tau/2)$. This trajectory has the
free parameters $\Omega$ and $\tau$, where the first describes how fast the
phase switches in imaginary time and the second how long it stays in the
shifted minimum before it returns. In the limit $\Omega\tau\rightarrow\infty$,
$\varphi_B$ becomes an exact saddle point of the dissipative action $S_\eta$.
Evaluating the whole action $S$ for this trajectory corresponds to first order
perturbation theory in the circuit action $S_c$. This approach leads to a
bounce action $S_B(\Omega,\tau)$ still depending on the two free parameters of
the bounce. To find the approximate saddle point, we need to extremize with respect to 
these parameters. We find two distinct regimes: the first one corresponds to
the regime found in Ref.~\onlinecite{Korshunov1987} that is valid as long as
$\Omega$ stays approximately constant. We denote the regime at
small bias current $j< (\zeta/2\eta)^{1/2}$ by (i). In this regime, we find
\begin{align}
\label{Eq.:KorshunovParameters}
  \tau^\text{(i)}=2\biggl(\frac{2\hbar\eta}{j E_J\omega_0^2}\biggr)^{1/3}\hspace{5pt} \text{and} \hspace{5pt} \Omega^\text{(i)}=\omega_0,
\end{align}  
resulting in the action 
\begin{align}
\label{Eq.:KorshunovAction}
  S_B^\text{(i)}=4\pi[2\eta+4\zeta-3(2\eta j^2\zeta^2)^{1/3}].
\end{align}

At elevated bias currents $(\zeta/2\eta)^{1/2}< j <j_\text{crit}\approx 0.2$,
we find a second novel regime in which the frequency $\Omega$ starts to decay
$\propto j^{-2}$. In this regime, we have to minimize in both parameters
$\Omega$ and $\tau$ , see App.~\ref{App.A} for more information.  The
resulting saddle point solution is given by
\begin{align}
\label{Eq.:NewActionParameters}
  \tau^\text{(ii)}=\frac{2}{j \Omega^\text{(ii)}} \quad\text{and}
  \quad\Omega^\text{(ii)}=\frac{E_J}{\hbar\eta j^2},
\end{align}
with
\begin{align}
\label{Eq.:BounceActionResult}
S^\text{(ii)}_{B}&=8\pi\eta(1-j^2).
\end{align}
As $j\lesssim 0.2$, the term $8\pi\eta$, which is the quasiparticle action
contribution of two infinitely separated instantons, always dominates. This is
in agreement with our assumption that the dissipative term approximately
determines the saddle point. If we exceed the critical current
$j_\text{crit}$, the extremum for $S_B(\Omega,\tau)$ is found at $\Omega=0$
and therefore the bounce of the dissipative action $S_\eta$ approaches the
constant solution $\varphi_0$ that stays in the minimum of the Josephson
potential. In Fig. \ref{Fig:ActionPlots}, we compare Eq. (\ref{Eq.:KorshunovAction}) and Eq. (\ref{Eq.:BounceActionResult}) to the value of $S_B(\Omega,\tau)$ at the saddle point that we obtained numerically.

\subsection{Bounce Prefactor and Result}

\begin{figure}
  \centering
\includegraphics[width=.9\linewidth]{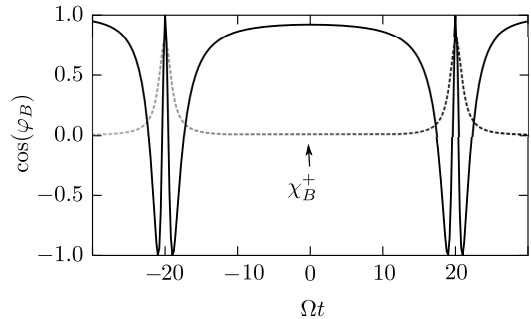}
\caption{\label{Fig:FluctuationPotPlot} The solid line shows the effective
potential $\cos(\varphi_B)$ of the circuit action $S_c$ for the fluctuations
around the bounce path (for $\Omega\tau=40$). The potential consists of two
double wells at $\pm\tau/2$. The dashed line corresponds to the even eigenmode
$\chi_{B}^+$ of the fluctuation operator. It is approximately given by a
superposition of the (shifted) instanton eigenmodes $\chi_0(t\pm\tau/2)$ where
the brighter part of the curve corresponds to $\chi_0(t+\tau/2)$ and the
darker part to $\chi_0(t-\tau/2)$.}
\end{figure}
To find the  $4\pi$-tunneling rate, the remaining task is to calculate the
prefactor $F_B$ that represents the quantum fluctuations on top of the bounce
path. The procedure is similar to the calculations for the instanton, however
with some complications added. First of all, we have to evaluate the
fluctuation operator at the bounce trajectory so that the eigenvalue equation
does not take the simple form (\ref{Eq.:FluctuationDissipation}).
We can approximate the exact fluctuation operator of $S_\eta$ by the form
\begin{align}
\label{Eq.:BounceEigenvalueEq}
& -\frac{2 \Omega}{1+\Omega^2 (t-\tau/2)^{2}} \biggl[ \chi_{B,n} (t)
  -\int \frac{dt'}{\pi} \frac{\Omega \,\chi_{B,n} (t')}{1+ \Omega^2
(t'-\tau/2)^{2}} \biggr] \nonumber\\
&-\frac{2 \Omega}{1+\Omega^2 (t+\tau/2)^{2}}\biggl[ \chi_{B,n} (t) -\int
\frac{dt'}{\pi} \frac{\Omega\,\chi_{B,n}(t')}{1+ \Omega^2 (t'+\tau/2)^{2}} \biggr]\nonumber\\
&+\int\!\mathcal{P} \frac{dt'}{\pi} \frac{1}{(t-t')} \frac{d
\chi_{B,n}(t')}{d t'} =\frac{\Lambda_{B,n}}{\eta} \chi_{B,n} (t)
\end{align}
valid for $\Omega\tau\to \infty$.  It corresponds to the instanton eigenvalue
equation \eqref{Eq.:FluctuationDissipation} with a potential at each positions
$\pm\tau/2$  of the constituting instantons. For large $\Omega\tau$, the
potential are well-separated, so that the eigenmodes are expected to be
superpositions of the instanton eigenmodes.

For example, for the low energy eigenvalues needed in the factor $A_{B,1}$, we
can make the ansatz of the even and odd superposition of the shifted instanton
zero modes
\begin{align}
\label{Eq.:ZeroModeSuperposition}
\chi^\pm_{B}=&\frac{1}{(N^\pm_{B})^{1/2}}(\chi_{0,+}\pm\chi_{0,-})\\
=&\frac{1}{(2\pi N^\pm_{B})^{1/2}}\biggl[\frac{\Omega^{1/2}}{1+ \Omega^2
(t+\tau)^{2}}\pm\frac{\Omega^{1/2}}{1+ \Omega^2 (t-\tau)^{2}}\biggr].
\nonumber
\end{align}
where $\chi_{0,\pm}=\chi_0(t\pm\tau/2)$ and the new normalization is given by
$N^\pm_{B}=[2\pm8/(4+\Omega^2\tau^2)]^{1/2}$. 

By comparing (\ref{Eq.:ZeroModeSuperposition}) with the derivative of the
bounce in respect to the imaginary time we see that the odd superposition
indeed corresponds to the real zero mode. This zero mode generates a shift of
the whole bounce trajectory in imaginary time. Moreover, approximately (up to
$\mathcal{O}[(\Omega\tau)^{-4}]$),  the even superposition is a zero mode of
the quasiparticle action, too. It is a so-called breathing mode and generates
a shift of the two instanton parts of the bounce in two different directions,
changing the size of the bounce. It can also be obtained by taking the
derivative of the bounce in respect to $\tau$.  Counting the numbers of nodes
we recognize that the zero mode $\chi_{B}^-$ has one node, while the even
mode $\chi_{B}^+$ has no nodes. Therefore, the even eigenvalue has to be
negative. For a negative eigenvalue the naive Gaussian fluctuation
approximation breaks down. However, it is this negative eigenvalue that gives
rise to the imaginary part of the energy that corresponds to the decay rate
\cite{Coleman1979}. 

The degeneracy between the even and odd mode is lifted if we perturbatively
consider the fluctuations of the circuit action. However, here we cannot take
the simple approach as in (\ref{Eq.:InstantonPerturbation1}) for the
instanton.  We encounter the problem that we do not know the eigenfunctions
accurately enough for this treatment. The eigenfunctions of the quasiparticle
action are a good approximation to the real eigenmodes away from the position
$\pm\tau/2$  of the instantons. However, close to these positions the
eigenmodes are subject to `fast' modulations that are not included in lowest
order perturbation theory. 

In Fig.~\ref{Fig:FluctuationPotPlot}, we show the Josephson potential for the
fluctuations around the bounce and the even mode $\chi_{B}^+$ to visualize
the problem. The eigenfunction $\chi_{B}^+$ plotted as the dashed line is
clearly not a ground state for the potential close to the points with $t=\pm
\tau/2$. The missing fast modulations are irrelevant for the quasiparticle
action but change the contribution by the Josephson potential already on the
order of $\zeta$. However, the splitting between the even and odd mode is of
the order $(\Omega\tau)^{-2}$ and thus we have to apply a modified procedure.

The idea is to   directly calculate the splitting $\Delta\Lambda_{B}$
between the two lowest eigenvalues $\Lambda_{B,0}$ and $\Lambda_{B,1}$ instead of
finding their absolute values. Knowing that $\Lambda_{B,1}=0$ for the exact
solution of the problem, we obtain $\Lambda_{B,0} = - \Delta\Lambda_{B}$. It
is possible, to calculate  $\Delta\Lambda_{B}$ without accurate knowledge
of  the wavefunctions close to the instanton position. For that we define
$T_\text{kin}=-(\hbar/8E_C) (\partial/\partial t)^2$, $V_\pm=E_J
\cos[\varphi_I(t\pm\tau/2)]/\hbar$, and $V_\text{pert}=V_0-V^+-V^-$ with
$V_0=E_J \cos(\varphi_B)$ the Josephson potential evaluated at the bounce. We
can rewrite the circuit fluctuation operator as
\begin{align}
\frac{\delta^2
S_c}{\delta\varphi^2}[\varphi_B]&=T_\text{kin}+V^++V^-+V_\text{pert}.
\end{align}
In the expression 
\begin{align}
\label{Eq.:EIgenvaluesBounceLow}
\Delta\Lambda_{B}&=\int\!dt
\Biggl[\chi_{B}^-\biggl(T_\text{kin}+V^++V^-+V_\text{pert}\biggr)\chi_{B}^-\nonumber\\
&\qquad-\chi_{B}^+\biggl(T_\text{kin}+V^++V^-+V_\text{pert}\biggr)\chi_{B}^+\Biggr]\nonumber\\
&=2\int \!dt \,\chi_{0,+}(V^-+V^++2V_\text{pert})\chi_{0,-}\nonumber\\
&=\frac{4E_J}{\hbar(\Omega\tau)^2}
\end{align}
for the first order perturbation, we make use of the fact that
$(T_\text{kin}+V^\pm)\chi_{0,\pm}=0$ for the zero mode. This removes the terms
$V^\pm\chi_{0,\pm}^2$ that are localized in the dangerous region around the
instanton position. Additionally, for the second equality, we have left out
terms proportional to $V^\mp\chi_{0,\pm}^2$ that are higher order in $\Omega
\tau$.

For the modes with more than one nodes ($n>1$), the accuracy of  the
conventional perturbation theory is sufficient. By using the odd superposition
of the shifted $\chi_{1,\pm}$ instanton eigenmodes, we can estimate the third
eigenvalue as $\Lambda_{B,2}=\hbar\Omega^2/16E_C$. The expression for the
normalization due to the zero mode reads $W_{B}= 2\pi
\hbar\Omega/E_C$. As a result, we obtain the prefactor
\begin{align}
A_{B,1}=2\sqrt{\frac{E_J}{\hbar \Omega}}\Omega\tau.
\end{align}

In order to calculate $A_{B,2}$, we still have to determine the higher
eigenvalues corresponding to $n\to\infty$. The high-energy eigenmodes are
still approximatley given by the eigenfunctions of the kinetic operator. We
obtain the corresponding eigenvalues by inserting the second variation
(\ref{Eq.:BounceEigenvalueEq}) of the bounce into the expression
\eqref{Eq.:UpperNEigenvalues}. We find the result
\begin{align}\label{eq:lambda_b}
  \Lambda_{B,2n-1}=\Lambda_{B,2n}=\frac{\hbar}{8 E_C}\bigl(\nu_n^2+\omega_0^2\bigr)+\eta|\nu_n|-2\eta\nu_1,
\end{align}
where the factor $2$ in the last term compared to
\eqref{Eq.:UpperNEigenvalues} originates from the fact that there are two
instantons contributing to the bounce. Plugging \eqref{eq:lambda_b} into
\eqref{Eq.:AFactorsGeneral2} yields (for $\eta \gg \zeta$)
\begin{align}\label{eq:lambda_b2}
A_{B,2}= (A_2)^2=\frac{\eta^4}{\zeta^4}.
\end{align}

With the results \eqref{eq:lambda_b}, \eqref{eq:lambda_b2}, and the ones in
Sec.~\ref{sec:bounce}, we are in the position to evaluate the decay rate for
the two regimes identified above. For low bias current
$j<(\zeta/2\eta)^{1/2}$, the rate is given by\cite{korshunov}
\begin{align}
\label{Eq.:RateResultKorsh}
\Gamma_{4\pi}^{(\text{i})}= \frac{\Delta_I^2\zeta^{-1/2}}{8\hbar^2}\biggl(\frac{2\hbar\eta}{j\omega_0^2E_J}\biggr)^{1/3}e^{12\pi(2\eta j^2\zeta^2)^{1/3}}.
\end{align}
For elevated currents with $(\zeta/2\eta)^{1/2}<j<0.2$, the decay rate is
given by
\begin{align}
\label{Eq.:RateResult}
\Gamma_{4\pi}^{(\text{ii})}=4\omega_0\frac{\eta^{7/2}}{\zeta^4}e^{-8\pi\eta(1-j^2)}.
\end{align}
The crossover from the result \eqref{Eq.:RateResultKorsh} to
\eqref{Eq.:RateResult} that we describe in more details below as well as the
decay rate  \eqref{Eq.:RateResult} at elevated bias current are the main
results of the present work.

\subsection{Regimes and Crossovers}

\begin{figure}
 \centering 
\includegraphics[width=.9\linewidth]{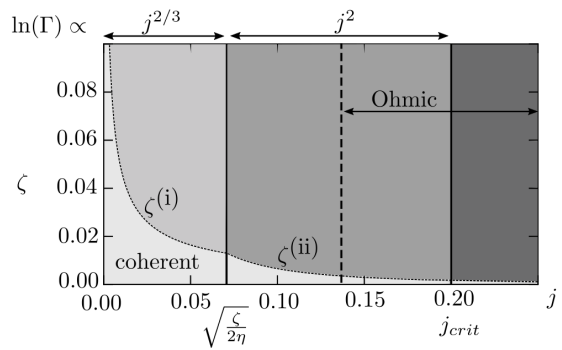}
\caption{\label{Fig:Regimes} The plot shows the crossovers between the
different regimes. The axis are given by the bias current $j$ and $\zeta$ (for
fixed $\eta/\zeta=100$). From light to dark color we go from the coherent
regime to the regime (i) with a scaling of the decay rate $\ln(\Gamma_{4\pi})\propto j^{2/3}$ followed
by the regime (ii) with $\ln(\Gamma_{4\pi})\propto j^2$ and finally end up with Ohmic
dissipation. The dashed black curve marks the crossover to the coherent
regime. For very small $\zeta$, there is no crossover to the $j^{2/3}$
regime. For approximately $j> 0.14$ (indicated by the dashed vertical line)
the rate of $2\pi$ decay generated by the Ohmic bounce becomes larger than the
rate of the $4\pi$-quasiparticle decay. However, the two processes can be
distinguished so that the quasiparticle decay can still be measured.   Above
$j_\text{crit}$, the bounce of the quasiparticle action approaches the
constant solution $\varphi_0=0$ and only the $2\pi$ process, where the
quasiparticle dissipation is approximately Ohmic, survives. For larger
$\zeta$ than shown on the figure, the coherent regime vanishes already for a small bias current, while the other crossover lines do not depend on $\zeta$.}
\end{figure}

In this section, we will discuss the crossovers between the regimes identified
above. Without bias current, the system forms bands due to dissipation
mediated coherent tunneling. We call this regime the `coherent regime', see
Fig.~\ref{Fig:Regimes}. The amplitude $\Delta_I/2$ then defines a tunneling
matrix element for a $4\pi$ phase slip. Increasing the bias current $j$, more
than a single state in the well separated by $4\pi$ becomes energetically
accessible and the coherent tunneling transforms into an incoherent
relaxation. A quantitative criterion for the crossover from the coherent to
the incoherent regime can be defined by
$(\tau^\text{(i,ii)})^2>\delta\tau^2\approx(\partial^2
S_B(\Omega,\tau)/\partial\tau^2)^{-1}$. This gives an estimate whether we can
treat the position $\tau$ of the bounce as a classical variable or whether
quantum fluctuations have to be taken into account.  As long as the quantum
fluctuations of $\tau$ are smaller than the optimal separation between the
instantons $\tau^\text{(i,ii)}$, the bounce and therefore incoherent tunneling
is an appropriate description. If the fluctuations in $\tau$ increase, the
system is more accurately described by a gas of individual instantons giving
rise to coherent tunneling elements.  Depending on the parameters, tuning $j$
up leads in general to a crossover of the action to the regime (i) with a
scaling of $\ln \Gamma$ $\propto j^{2/3}$ and then to the regime (ii) with a
scaling $\propto j^2$. However, for $\zeta<0.012$, the regime (i) is never
realized and the system directly crosses over from the coherent regime to the
regime (ii). From the crossover criterion above, we obtain the approximate
expressions for the crossover (at fixed $\eta/\zeta$)
\begin{align}
\zeta^\text{(i)}=\frac{1}{24\pi(10 +2\eta/\zeta)^{1/3}j^{2/3}}
\end{align}
in the regime $j<(\zeta/2\eta)^{1/2}$ and
\begin{align}
\zeta^\text{(ii)}=\frac{\eta/\zeta}{\pi(24 +(48\eta/\zeta-80)j^{2})}
\end{align}
in the regime $(\zeta/2\eta)^{1/2}<j<0.2$. At $j\approx0.14$ the rate of
$2\pi$ phase slips $\Gamma_{2\pi}$ generated by the Ohmic bounce solution is
of the same order as the quasiparticle decay $\Gamma_{4\pi}$. However, the two
processes are physically distinguishable so that they can be individually
measured; see also below. At a bias current $j$ above $j_\text{crit}$, the
bounce connecting two minima separated by $4\pi$ vanishes such that only
Ohmic dissipation is present in this regime.

\section{Measurement}
\label{Sec.:Measurement}

As demonstrated above, Josephson junctions with strong quasiparticle
dissipation admit many interesting properties that can be subject of an
experimental investigation. The simplest approach to observe effects of the
special (non-linear) form of the dissipation due to quasiparticle tunneling is
to measure the incoherent tunneling. Measuring the coherent tunneling directly
is challenging due to the small bandwidth exponentially suppressed in $\eta$
without any additional tuning parameter. Therefore, we propose to measure
paired phase-slip events and compare the resulting rates to the expressions
(\ref{Eq.:RateResultKorsh}) or (\ref{Eq.:RateResult}).

The key idea for the experimental observation of the paired phase slips is to
raise the bias as much as possible, \textit{i.e.}, smaller than
$j_\text{crit}$ but still in its vicinity, in order to increase the rate of
paired phase slips.  An important requirement for the experimental setup in
order to be able to operate at elevated bias current is the possibility to
distinguish between double and single phase slips. The reason is that at
elevated bias current, the rate of unpaired $2\pi$ phase slips can already
dominate the rate of paired $4\pi$ phase slips. Additionally, even if we can
distinguish between the two processes, we need to make sure that the $4\pi$
process can be uniquely associated with the periodic quasiparticle tunneling
while the $2\pi$ process is caused solely by the conventional Ohmic tunneling.
The latter process does not necessarily end up in the nearest minimum. If the
momentum, \textit{i.e.}, the kinetic energy, of the phase difference is to large
it may not be retrapped after the tunneling but it can classically go on over
the next potential hill and end up in the following minimum. Especially after
the point at which the Ohmic tunneling rate $\Gamma_{2\pi}$ exceeds the
quasiparticle rate $\Gamma_{4\pi}^{(i)}$ or $\Gamma_{4\pi,2}^{(ii)}$
respectively it is not possible to distinguish between the two processes
anymore. Therefore, it is best to keep the capacitance $C$ small so that the
dissipation always brings the Ohmic phase slips to rest in the next minimum.
Additional it is advantageous to use small $\zeta\ll1$ because it allows to
consider systems with smaller $\eta$ without making the ratio $\eta/\zeta$ too
large. Smaller $\eta$ then keeps the exponential suppression of the phase-slip
rate low.
 
An approach that can satisfy the above  requirements is to include the
Josephson junction into a loop with inductance $L$ or alternatively build an
asymmetric SQUID so that one Josephson junction serves as an inductance; see
Ref.~\onlinecite{Belkin2011,Belkin2015} for a recent experimental setup. With
a magnetic bias, it is possible to add an external flux $\varphi_{\text{ex}}$
in the loop that takes the role of the bias current.  Placing the circuit in a
transmission line, the number of flux quanta in the loop can be measured
non-destructively by a flux dependent shift of the transmission phase of the
input and output signal into the transmission line. This flux dependent shift
in the transmission phase directly indicates when a $2\pi$ or a $4\pi$-event
has happened. By recording these events over a given measurement time, the
resulting rates can be compared with the results (\ref{Eq.:RateResultKorsh})
or (\ref{Eq.:RateResult}). Theoretically, the setup corresponds to introducing
the additional term $S_L=\int\!dt\,
\phi_0^2(\varphi-\varphi_\text{ex})^2/(8\pi^2 L)$ to the circuit action $S_c$
with induction $L$. In this setup, the bias current is given by the term
linear to $\varphi$ with $I=\hbar\phi_0\varphi_\text{ex}/4\pi^2L$. The
additional quadratic contribution $\propto\varphi^2$ simply changes the bias
current according to $j\mapsto j-\hbar\phi_0^2/\pi L E_J$. This takes care of
the fact that a quadratic potential needs already an external flux of
$\varphi_\text{ex}=4\pi$ until the minimum at $\varphi=0$ becomes unstable for
the Korshunov decay channel while without the quadratic confinement an
infinitesimal bias is already enough to render the minimum unstable. A similar
setup has been used in Ref.~\onlinecite{Belkin2015} to measure the
interference between phase slips in two parallel nanowires. It indicates that
the quasiparticle tunneling in nanowires is strong and therefore these wires
are a potential candidate for such an experimental setup. Note that there are
also other potential experimental probes that can detect changes in flux. For
example, flux dependent absorption process could be used to measure the rate
of paired phase slips.\cite{Chiodi2011,Dassonneville2013}

\section{Conclusion}
\label{Sec.:Conclusion}

In conclusion, we have shown that the coherent dissipation due to
quasiparticle tunneling over a Josephson junction in a superconductor can be
probed by the measurement of 4$\pi$ phase-slip events. These $4\pi$-phase
slips are caused by Korshunov instantons probing the specifics of the
nonlinear dissipation due to quasiparticles. We have identified a novel regime
at elevated bias current that leads to a substantially increased rate of
4$\pi$ phase slips. This is important as the low rate is one of the main
reasons why paired phase slips are challenging to measure. We have discussed
the different crossovers between the coherent regime and the incoherent
regimes. In addition, we have proposed a measurement scheme for the detection of
the paired phase slips; fixing the bias current slightly below a critical
current $j_{\text{crit}}\approx0.2$ and working with a small capacitance $C$,
corresponding to a large charging energy, offers the best chance to observe
paired phase slips due to the increased rates. We hope that our analysis helps
to guide the experimental effort to directly observe Korshunov instantons as
paired phase slips of the superconducting phase.

\section{Acknowledgments} 

The authors acknowledge support from the Deutsche Forschungsgemeinschaft (DFG)
under grant HA 7084/2-1.

\appendix

\section{Extremizing the action in different regimes}
\label{App.A}
\begin{figure}
 \centering 
\includegraphics[width=.9\linewidth]{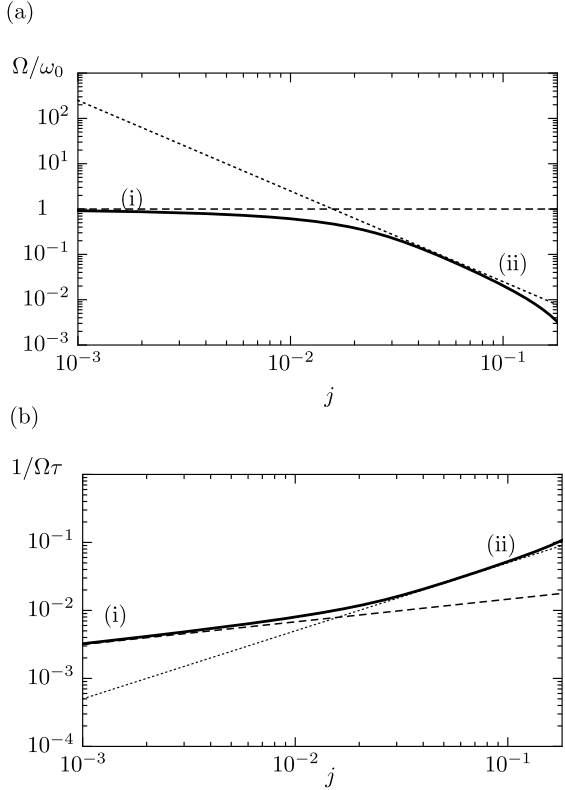}
\caption{\label{Fig:ExtremePlots} The figure shows double logarithmic plots of
the extremal parameters $\Omega/\omega_0$ (a) and $1/\Omega\tau$ (b) as a
function of the dimensionless bias current $j$ for the parameters $\eta=2$ and
$\zeta=10^{-3}$. The bold black lines correspond to the result that is
obtained by simply extremizing the whole action (\ref{Eq.:ExpandedAction}) in
respect to both parameters  $\Omega$  and $\tau$ numerically. The dashed line
represents regime (i), where $\Omega=\omega_0$ is constant. The dotted line
represents regime (ii) where $\Omega$ decays to zero.}
\end{figure}
In this appendix, we provide details for the extremizing the action $S_B$ to
find the regime (i) corresponding to equations (\ref{Eq.:KorshunovParameters})
and (\ref{Eq.:KorshunovAction}) and regime (ii) corresponding to equations
(\ref{Eq.:NewActionParameters}) and (\ref{Eq.:BounceActionResult}). The action
$S_B$ evaluated at the bounce trajectory and consistently expanded for large
$\Omega\tau$ up to second order reads
\begin{align}
\label{Eq.:ExpandedAction}
S_B&\approx
8\pi(\eta+E_J/\hbar\Omega+\hbar\Omega/8E_C)-4\pi E_J \tau j/\hbar\nonumber\\
&\quad-\frac{32\pi(\eta+4 E_J/\hbar \Omega+\hbar\Omega/8E_C)}{\Omega^2\tau^2}.
\end{align}
In Fig.~\ref{Fig:ExtremePlots}, we show an example of the resulting optimal
parameters calculated by a numerical optimization of the action above with
respect to $\Omega$ and $\tau$. It clearly shows two distinct regimes with
different power law behaviors. The first regime corresponds to regime (i) with
a constant $\Omega$ while the second regime corresponds to (ii) with decaying
$\Omega$.  The crossover is numerically found to be at
$j_c\simeq(\zeta/2\eta)^{1/2}$, see below.

An analytic expression valid in the regime (i) can be found by assuming
$\Omega=\omega_0$ and optimizing (\ref{Eq.:ExpandedAction}) with respect to
the single parameter $\tau$. In this case, only the two last terms in
\eqref{Eq.:ExpandedAction} contribute. This yields
Eqs.~(\ref{Eq.:KorshunovParameters}) and (\ref{Eq.:KorshunovAction}) for the
optimal point.

By increasing the bias current, the assumption $\Omega=\omega_0$ fails to hold
as the inverse size of the instanton $\Omega$ starts to decline with raising
bias current $j$. As a result the term $8\pi  E_J/\hbar\Omega$ starts to
become relevant. The point at which this happens can be estimated by comparing
it to one of the two last terms, e.g., $E_J/\hbar\omega_0 \simeq E_J\tau
j_c/\hbar$. With $\tau \simeq (\hbar \eta/j E_J \omega_0^2)^{1/3}$ [from
\eqref{Eq.:KorshunovParameters}], we obtain the estimate for the crossover
current $j_c\simeq (\zeta/\eta)^{1/2}$ as before.

So for $j \gg j_c$, the parameters $\tau$ and $\Omega$ in the action have to
be simultaneously optimized. Not all terms of the action
\eqref{Eq.:ExpandedAction} are relevant. In the first term, we can neglect the
term proportional to $\Omega$ as $\Omega \ll \omega_0$. In the last term, only
the term proportional to $\eta$ is relevant as $\eta \gg \zeta$. Thus, the
effective action in the regime (ii) reads
\begin{align}
\label{Eq.:ActionMostImportantTerms}
S_B\approx8\pi(\eta+E_J/\hbar\Omega)-4\pi E_J \tau j/\hbar-32\pi\eta/\Omega^2\tau^2.
\end{align}
Extremizing this action with respect to the parameters $\Omega$ and $\tau$ is
straightforward and leads to the results of
Eq.~(\ref{Eq.:NewActionParameters}).  Inserting the optimized parameters into
Eq.~(\ref{Eq.:ActionMostImportantTerms}) yields the simple expression for the
action
\begin{align}
S_B\approx 8\pi\eta(1-j^2),
\end{align}
that is equivalent to (\ref{Eq.:BounceActionResult}). For bias currents $j>0.1$, the accuracy of (\ref{Eq.:BounceActionResult}) can be increased by including small corrections to the $j^2$-dependence with first order perturbation theory. This corresponds to inserting the optimized values $\Omega$ and $\tau$ from Eq.~(\ref{Eq.:NewActionParameters}) into the full action (\ref{Eq.:ExpandedAction}).

\end{document}